\documentclass[structabstract]{aa}
\usepackage{graphicx}
\usepackage{txfonts}
\begin{document}
   \title{ Notes on the disentangling of spectra}

   \subtitle{I. Enhancement in precision}

   \author{P. Hadrava
          }

   \institute{Astronomical~Institute, Academy~of~Sciences,
              Bo\v{c}n\'{\i}~II~1401, CZ~-~141~31~Praha~4,
              Czech~Republic\\
              \email{had@sunstel.asu.cas.cz}
             }

   \date{Received 15 August 2008 / Accepted 16 October 2008}


  \abstract
  {The technique of disentangling has been applied to numerous
  high-precision studies of spectroscopic binaries and multiple
  stars. Although, its possibilities have not yet been fully understood
  and exploited.}
  {Theoretical background aspects of the method, its latest
  improvements and hints for its use in practice are explained
  in this series of papers.}
  {In this first paper of the series, we discuss spectral-resolution
  limitations due to a discrete representation of the observed spectra
  and introduce a new method how to achieve a precision higher
  than the step of input-data binning.}
  {Based on this principle, the latest version of the KOREL code for
  Fourier disentangling achieves an increase in precision for an order
  of magnitude.}
   {}


   \maketitle
%

\section{Introduction}
Disentangling spectra of binary and multiple stars enables us
to determine efficiently the orbital parameters and simultaneously
to separate the spectra of the component stars.
This numerical technique performed in either the wavelength domain
(Simon and Sturm 1994) or its Fourier image (Hadrava 1995)
has been applied successfully in numerous
studies of individual stellar systems. However, some users failed in
their attempts or were unable to take full advantage of the method,
partly due to a misunderstanding of its principles. A review of
the Fourier disentangling has been provided by Hadrava (2004) together
with the release in 2004 of the author's code KOREL, but, regarding
new improvements of the method, this review is already out of date.
The purpose of the present series of papers is to explain some
common mistakes, provide practical hints for using the method,
and present its new developments.

 In this paper, consequences of the discretization of the observed
spectra are discussed in Sect.~\ref{Sampl}. A new method for
enhancement of the spectral resolution in disentangling spectra
is introduced in Sect.~\ref{Enhance}. Results and their implications
are briefly summarized in Sect.~\ref{Concl}.


\section{Sampling of the input spectra}\label{Sampl}
In their study of disentangling, Hensberge et al. (2008)
speculated about ``expense" at which the computational efficiency
of the spectral method dealing with the Fourier image surpasses
the method of singular-value decomposition in the wavelength
domain. They suggested that, among other things, it may be the need
of having the input
observed spectra sampled on a common grid equidistant in the logarithmic
wavelength scale
\begin{equation}\label{scale}
 x=c\,{\rm ln}\lambda/\lambda_{0}\; ,
\end{equation}
where $\lambda_{0}$ is an arbitrarily chosen reference wavelength.
It should be noted that the same assumption is commonly imposed
on the solution in the wavelength domain as well, as it has been
described by Simon and Sturm (1994) and it is also obvious from
the explanation in Fig.~1 of Hensberge et al. (2008) or their example
in Appendix of the same paper.

 The uniform sampling of input data simplifies the solution, but this
assumption may be avoided in both methods of disentangling. In the
Fourier view, it is obvious that the Fourier transforms $\tilde{I}(y)$
of the observed spectra $I(x)$ in chosen (equidistant) sampling frequencies
$y_{k}$ may be calculated directly according to the definition
\begin{equation}\label{FT}
 \tilde{I}(y_{k})=\int I(x)\exp(iy_{k}x)dx
\end{equation}
from any original (even non-equidistant) binning $x_{l}$ if the function
$I(x)$ is suitably interpolated, e.g. by the simple linear formula
\begin{eqnarray}\label{Interpol}
 I(x)&=&I(x_{l})\frac{x_{l+1}-x}{x_{l+1}-x_{l}}
      +I(x_{l+1})\frac{x-x_{l}}{x_{l+1}-x_{l}}\\
     && \hspace*{47mm}{\rm for}\;x\in(x_{l},x_{l+1})\; .\nonumber
\end{eqnarray}
The common practice of interpolating $I(x)$ first to the equidistant grid
points and then using the Fast Fourier Transform saves the computer time
(at some expense of accuracy), but is not inevitable in Fourier
disentangling.

In the wavelength-domain solution,
the single off-diagonal matrices ({\bf N} in notation of Simon and
Sturm 1994) shifting the spectra of component stars to their appropriate
positions in individual exposures may be replaced by wider band matrices
if the observed spectra are not sampled in the same equidistant set
of the logarithmic wavelengths into which the component spectra are to
be separated (Simon and Sturm 1994, p. 287: ``The submatrices of {\bf M},
{\bf N}$_{Ai}$ and {\bf N}$_{Bi}$, are rectangular band matrices with
a bandwidth depending on the differences in dispersion of the wavelength
scales of {\bf c} and {\bf x}").
A simple possibility is to use a matrix {\bf N} with two
non-zero elements in each column given by Eq.~(\ref{Interpol}).
This is, however, again equivalent to a suitable
resampling, and a subsequent solution in the convenient representation
of the observed spectra. An additional significance of using bandwith
matrices to refine the model is discussed in the next Section.

 Although any resampling implies some smoothing of the input signal and thus
a loss of information in the high-frequency modes, it is inevitable provided
the observations are not directly obtained in the required data bins.
The question is therefore not about performing a resampling, but how
it can be performed best.
This problem, which is common to any method of disentangling, is related
to the more general task of optimal data-processing of observed spectra
(cf. Hensberge 2004) and its aspects in disentangling will be studied in
detail elsewhere.

 A consequence of discretizing observed spectra is limitation of
the accuracy at which the radial velocities are determined. Until
now, a common practice in both the wavelength-domain and Fourier
disentangling was to round the expected Doppler shift to an integer multiple
of the radial-velocity step. This limited resolution of Doppler shifts in
the individual spectra also limits the precision of the disentangled orbital
elements and the sharpness of the separated spectra of component
stars. A straightforward means of improvement appears to be a choice of
a smaller sampling step. However, the resolution is limited by the detectors
in any case. On the other hand, it is intuitively evident that if we have
a set of spectra with mutually shifted sampling, we can also reconstruct
details on a sub-pixel scale. For instance, if a very narrow line with
a sub-pixel width moves from a given pixel in some exposures to the
neighbouring pixel in other exposures, then its position may be found
precisely from the time and its width from the duration of this transition.
It is thus worth investigating the limits of resolution in details.


\section{An increase of spectral resolution}\label{Enhance}
In my spectral method, the shift in the spectra $I(x)$ of each component
(which we wish to separate from the observed superpositions),
in the logarithmic wavelength scale $x$ defined by Eq.~(\ref{scale})
for a value $v$ of the radial velocity, is given by the convolution
of the spectra with the shifted Dirac delta-function $\delta(x-v)$,
\begin{equation}\label{Kor1}
 I'(x)=I(x-v)=I(x)\ast\delta(x-v)\; ,
\end{equation}
which implies, in the Fourier transform ($x\rightarrow y$), a multiplication
by a function $\exp(iyv)$,
\begin{equation}\label{FKor1}
 \tilde{I}'(y)=\tilde{I}(y)\exp(iyv)
\end{equation}
(cf. $I_{j}$ in Eqs.~(1) and (2) of Hadrava 1995).
This simple exponential function can be evaluated precisely at each
frequency $y$. However, due to the limited number $N$ of the modes taken
into account, its inverse Fourier transform will generally produce a wider
peak with some ghosts on its sides resembling interference fringes. Only in
the special case of $v$ being an integer multiple of the grid step,
the period of function $\exp(iyv)$ is in resonance with the interval
length in $y$-representation and a sharp shifted $\delta$-function
coinciding with a grid point of the $x$-representation can be reproduced.
For that reason the radial velocity was rounded to the nearest grid point
in the Fourier disentangling also and it explains why the radial velocities
or their residuals calculated by the original KOREL-code were quantized
depending on the radial-velocity step.

   \begin{figure}
   \centering
   \includegraphics[width=9cm]{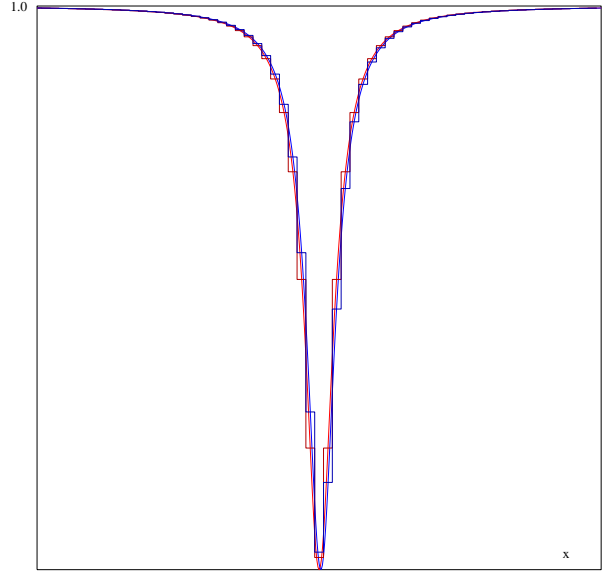}
   \caption{ Discretization of a Lorentzian profile $I(x)$ (the smooth thick
 line) centered on the pixel position should yield a symmetric distribution
 of counts $D[I](x)$ in neighbouring bins (the thick step function).
 A slightly shifted profile $I'(x)$ (for 0.2 pixel-width in this figure --
 see the thin lines) results in an asymmetry of the counts $D[I'](x)$, which
 in turn enables us to determine the line position at a precision below
 the pixel width.}
 \label{obr1x}
   \end{figure}

 However, owing to the resolution in the digitalized values of intensity
read from individual detector pixels, the position of spectral lines
wider than the sampling step can be deduced with an accuracy exceeding
the step width (cf. Fig~\ref{obr1x}). Alternatively to a convolution
with the shifted $\delta$-function, a shift of a spectrum $I(x)$ for
value $v$ can be expressed as a Taylor expansion
\begin{equation}\label{Taylor}
 I(x-v)=\sum_{j=0}^{\infty}\frac{1}{j!}I^{(j)}(x)(-v)^{j}\; ,
\end{equation}
which usually converges rapidly for small values of $v$. In a discrete
equidistant representation $x_{k}$ with the step $\Delta_{x}$, the first
two derivatives can be approximated by finite differences
\begin{eqnarray}\label{Taylor1}
 I^{(1)}(x_{k})&\simeq&\frac{1}{2\Delta_{x}}(I(x_{k+1})-I(x_{k-1}))\; ,\\
 I^{(2)}(x_{k})&\simeq&
  \frac{1}{\Delta_{x}^{2}}(I(x_{k+1})-2I(x_{k})+I(x_{k-1}))\; .\label{Taylor1a}
\end{eqnarray}
Therefore in the vicinity of the grid point $x_{k}$ a small shift of $I$
can be expressed in terms of values in this and the two neighbouring points as
\begin{eqnarray}\nonumber
 I(x_{k}-v)&\simeq& I(x_{k})-I^{(1)}(x_{k})v
 +\frac{1}{2}I^{(2)}(x_{k})v^{2}+o(v^{3})\simeq\\
 &\simeq& I(x_{k})-\frac{v}{2\Delta_{x}}(I(x_{k+1})-I(x_{k-1}))+ \label{Taylor2}\\
&&\hspace*{-10mm} +\frac{v^{2}}{2\Delta_{x}^{2}}(I(x_{k+1})-2I(x_{k})+I(x_{k-1}))
 +o(v^{3}) \; .\nonumber
\end{eqnarray}
This implies that the operator $\delta(x-v)$ of the shift is approximated
by
\begin{eqnarray}\label{Taylor3}
 \delta(x-v)&\simeq& \delta(x)
 -\frac{v}{2\Delta_{x}}(\delta(x+\Delta_{x})-\delta(x-\Delta_{x}))+\\
&&\hspace*{-10mm} +\frac{v^{2}}{2\Delta_{x}^{2}}(\delta(x+\Delta_{x})
 -2\delta(x)+\delta(x-\Delta_{x})) +o(v^{3}) \nonumber
\end{eqnarray}
and its Fourier transform
\begin{eqnarray}\nonumber
 \exp(iyv) &\simeq &1-\frac{v}{2\Delta_{x}}(\exp(-iy\Delta_{x})-\exp(iy\Delta_{x}))+\\
 &&\hspace*{-10mm}  +\frac{v^{2}}{2\Delta_{x}^{2}}(\exp(-iy\Delta_{x})
 -2+\exp(iy\Delta_{x})) +o(v^{3})=\nonumber \\
 &\hspace*{-19mm}=&\hspace*{-10mm}1+\frac{iv}{\Delta_{x}}\sin(y\Delta_{x})
 +\frac{v^{2}}{\Delta_{x}^{2}}(\cos(y\Delta_{x})-1) +o(v^{3})\label{Taylor4}\; .
\end{eqnarray}

   \begin{figure}
   \centering
   \includegraphics[width=9cm]{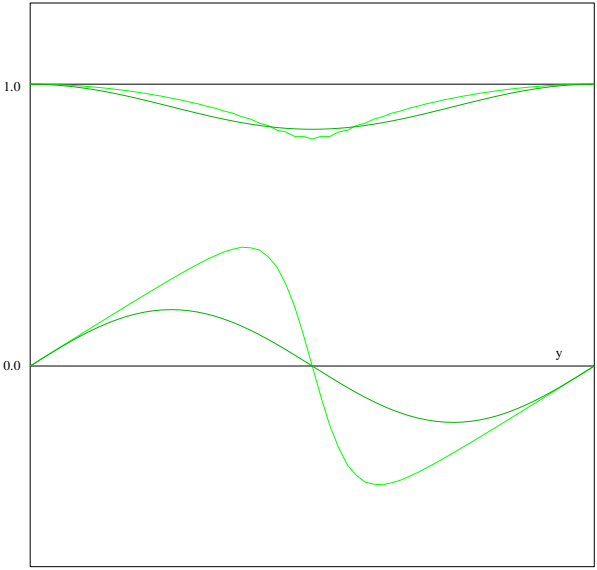}
   \caption{
Fourier
 transform of the discretized shifted profile $D[I']$ corresponds to the Fourier
 transform of the centered profile $D[I]$ multiplied by a correction, the real and
 imaginary parts of which are drawn by the upper and lower (distorted
 sinusoidal) thick lines, resp., while their approximations (\ref{Taylor4}) are
 drawn by the thin sinusoidal lines.
   }
 \label{obr2x}
   \end{figure}

 It can be seen in Fig.~\ref{obr2x} that a ratio of the two Fourier transforms
of mutually shifted profiles is approximated well by this simple sinusoid
for small values of $y$, while for the higher-frequency modes (drawn closer
to the middle of the figure) higher harmonics contribute significantly. This
is an obvious consequence of the fact that the approximations (\ref{Taylor1})
and (\ref{Taylor1a}) of the derivatives are more accurate for the lower
modes, which do not change significantly on the scale of $\Delta_{x}$.
The exact shape of the shift operator depends on the spectrum to be
shifted, unless $v$ is an integer multiple of $\Delta_x$. It means
that the value of $v$ cannot be disentangled with unlimited precision
from roughly sampled, unknown spectra. However, already the application
of the correction (\ref{Taylor4}) improves the precision of the
disentangling significantly, and the accuracy could be even higher for
disentangling constrained by a template spectrum.

 The above described procedure of reconstructing component spectra
from a large set of observations should not be confused with a simple
interpolation (given e.g. by Eq.~(\ref{Interpol})) in a single observation
or between grid points of some theoretical models. For instance,
in model atmospheres, the dependent variable (e.g. the specific intensity)
is calculated usually for chosen exact values of the independent variable
(wavelength) from which it can be interpolated to other values or integrated
over some regions of the independent variable. On the other hand, in true
observations, the values read at individual detector pixels are the quantities
integrated over some interval of the wavelength, which provide some
constraint only on the inner distribution. Without any additional information,
these values may be used in a single exposure as an estimate of the variable
for the middle of the interval, while closer to its edges the value
interpolated between the neighbouring bins is more appropriate. However, for a set of exposures,
the information can be combined to reveal partly also the subpixel structure,
or, using the above described procedure, to find subpixel mutual shifts
between the exposures.

 As an example, we show a result of disentangling of simulated data.
Twenty spectra uniformly covering one period (which is taken to be
a unit of time)
of a double-line binary on circular orbit with chosen radial-velocity
semi-amplitudes $K_{1}=50$km/s for the primary, and $K_{2}=100$km/s for
the secondary were calculated.
For each component, one line with a Lorentzian profile (with central depths
0.3 and 0.2 of the common continuum and half-widths equal to 30 and 40 km/s
for the primary and secondary, respectively) was included. A pseudo-random noise
scaled to amplitudes $n=$ 0\%, 0.5\%, 1.\%, or 2.\% of the continuum level was
added. The spectra were sampled by integrating in bins of width corresponding
to 10 km/s. The results of disentangling obtained using the KOREL code in its
old version (KOREL04 released by the author in 2004) and in its new version
(KOREL08) of enhanced precision, are compared in Table~\ref{example}.
In this Table, $S$ denotes the integrated square of spectra residuals, and
$\Delta T_{0}$ is the difference in units of the period between the solved
epoch of periastron (defined by fixed periastron longitude) and its true value
$T_{0}=0$ chosen for the simulation. Similarly, $\Delta K_{1}$ are the
differences between the calculated and true radial-velocity semi-amplitudes
of the primary and $\Delta q$ for the mass ratio
($q=M_{2}/M_{1}=K_{1}/K_{2}=0.5$).

 It can be seen from the results that the squares $S$ of the residuals
consist of a part approximately proportional to the square $n^{2}$ of
the noise, as can be supposed, but also an other additive, almost
constant part, which is comparable to the 1\% noise in the solution with
the classical KOREL04, but is suppressed for at least two orders in
the super-resolution KOREL08. This part is obviously due to the discrepancies
between sampling of the component spectra in different exposures shifted
by a non-integer multiple of the sampling step. This contribution depends
on the shape of the spectrum and its importance on the level of the noise.
This explains why in preliminary applications to real data the new method
yielded significantly superior results in some cases, but only a negligible
improvement in other cases.

 Similarly, the errors in orbital parameters have a part that increases with
the noise and a noise-independent part, which is significantly smaller in
the solution based on the new KOREL08.


\begin{table}
 \caption[]{Comparison of the old and new KOREL solutions.}
  \label{example}
 $$
   \begin{array}{llll}
     \hline
     \noalign{\smallskip}
     {\rm Parameter} & n [\%] &{\rm KOREL04} & {\rm KOREL08} \\
      \noalign{\smallskip}
      \hline
      \noalign{\smallskip}
      S                  & 0   & 3.61 & 0.012 \\
                         & 0.5 & 4.98 & 1.22 \\
                         & 1.  & 8.76 & 4.85 \\
                         & 2.  & 23.59 & 19.33 \\
      \noalign{\smallskip}
      \hline
      \noalign{\smallskip}
      \Delta T_{0}\; [Period] & 0   & 4.9 \times10^{-4} & 4. \times10^{-9} \\
                         & 0.5 & 4.9 \times10^{-4} & 6.6 \times10^{-6} \\
                         & 1.  & 4.9 \times10^{-4} & 2.3 \times10^{-4} \\
                         & 2.  & 7.2 \times10^{-4} & 5.0 \times10^{-4} \\
      \noalign{\smallskip}
      \hline
      \noalign{\smallskip}
      \Delta K_{1}\; [km/s] & 0 & 0.49 & 0.07 \\
                         & 0.5 & 0.49 & 0.03 \\
                         & 1.  & 0.49 & 0.06 \\
                         & 2.  & 0.33 & 0.18 \\
      \noalign{\smallskip}
      \hline
      \noalign{\smallskip}
      \Delta q\;         & 0   & 0.010 & 0.0005 \\
                         & 0.5 & 0.010 & 0.0002 \\
                         & 1.  & 0.010 & 0.0014 \\
                         & 2.  & 0.013 & 0.0040 \\
      \noalign{\smallskip}
      \hline
   \end{array}
     $$
\end{table}


\section{Conclusions}\label{Concl}
The correction provided in Eq.~(\ref{Taylor4}) for the residual
part of the radial velocities over an integer multiple of the sampling
step improves the Fourier disentangling significantly.
With this result and other improvements completed by the author to recent
versions of the KOREL code, the version of 2004 is no longer supported,
and we recommend using for true applications the version of 2008.

 Due to the equivalence of Eqs.~(\ref{Kor1}) and (\ref{FKor1}),
the wavelength-domain solution could be improved similarly if
the single off-diagonal matrix {\bf N} would be replaced by a three-
(off-)diagonal matrix (\ref{Taylor3}), or even by a more complicated matrix,
if it should also include an interpolation from a non-uniform
sampling of the input data. A possibility for using band-matrices was
mentioned by Simon and Sturm (1994), and in more detail
explained by Sturm (1994) (cf. also Hensberge et al., 2008).

 The method described here could be improved to achieve an even higher
precision for known component spectra (i.e. for the constrained
disentangling), for which the higher harmonics of the approximation
given by Eq.~(\ref{Taylor4}) could be estimated. Analogous numerical
refinement either in direct or Fourier space could be useful also in
other methods in spectroscopy (e.g. in methods using the broadening
function) as well as in data processing in other fields of astrophysics.

\begin{acknowledgements}
 This work has been completed in the framework of the Center for Theoretical
Astrophysics (ref.~LC06014) with a support of grant GA\v{C}R 202/06/0041.
The comments by the referee Sa\v{s}a Iliji\'{c} are highly appreciated.
\end{acknowledgements}

\end{document}